\documentstyle[prl,aps]{revtex}
\begin{document}
\twocolumn[
\draft
\title{Ground states with cluster structures in a frustrated 
Heisenberg chain}
\author{K Takano$\dagger$, K Kubo$\ddagger$ and 
H Sakamoto$\ddagger$} 
\address{$\dagger$Toyota Technological Institute, Tenpaku-ku, 
Nagoya 468, Japan \\
$\ddagger$Institute of Physics, Tsukuba University, Tsukuba 305, 
Japan} 
\hspace*{6cm}  Received
%
\maketitle
\widetext
\begin{quote}
   We examine the ground state of a Heisenberg model with arbitrary 
spin $S$ on a one-dimensional lattice composed of diamond-shaped 
units. 
   A unit includes two types of antiferromagnetic exchange interactions 
which frustrate each other.
   The system undergoes phase changes when the ratio $\lambda$ 
between the exchange parameters varies.
   In some phases, strong frustration leads to larger local structures 
or {\it clusters} of spins than a dimer. 
   We prove for arbitrary $S$ 
that there exists a phase with four-spin cluster states, which was 
previously found numerically for a special value of $\lambda$ 
in the $S=1/2$ case. 
   For $S=1/2$ we show that there are three ground state phases 
and determine their boundaries.
\end{quote}
] 
\narrowtext
   Effects of frustration in quantum antiferromagnets are of great 
current interest in solid state physics. In a classical system, 
strong frustration obstructs a simple antiferromagnetic (AF) 
ordering and produces another magnetic order characteristic of 
each system, e.g. the 120$^\circ$ structure or a spiral \cite{kaplan}. 
   In a quantum system, the interplay of quantum fluctuations and 
frustration makes the situation more complicated. 
   There may appear an exotic ground state which does not have 
magnetic order and has no classical analog. 
   A typical example is the complete dimer state in the 
Majumdar-Ghosh (MG) model \cite{majumdar-g}. 

   There are two types of frustrated quantum spin systems. 
   One is a system the classical version of which has a locally 
stable spin configuration in the ground state; i.e. any local 
deformation for a spin configuration always raises the energy.
   We say that such a system is {\it elastic}.  The MG model, the model 
with linearly decreasing AF interactions \cite{takano,linear} and the 
AF Heisenberg model on the triangular lattice are of this type.
   For the other type, the classical version of a system has ground-state 
spin configurations which can be locally deformed without raising the 
energy. Then the set of these configurations is a manifold with 
dimensions proportional to the system size.
   We say that a system of this type is {\it floppy}. The AF Heisenberg 
models on the $\Delta$ chain \cite{delta}, the double chain with 
diagonal interaction \cite{double} and the kagom\'e lattice 
\cite{kagome} are of this type. Some other floppy spin systems 
are also seen in  \cite{takano}. 

   Specially interesting is floppy systems. For example, it is argued 
that the kagom\'e antiferromagnet has a mysterious peak in a low 
temperature part of the specific heat \cite{kagome}. In spite of quite a 
few theoretical studies on this system, its quantum ground state and 
the low temperature thermodynamic properties are hardly clarified. 
   The difficulty of the problem originates from the floppiness. 
In a classical floppy system, the configuration of a local set of spins 
can be deformed without affecting the other part.
In the corresponding quantum system the local set may form a nearly 
closed state, or a {\it cluster}. A part surrounded by such clusters 
forms another cluster. Thus the total wave function becomes 
approximately of a direct product form. 
   To confirm this picture, it is important to examine a simple floppy 
model in which the structure of the ground state is clearly seen. 

   The singlet dimer is the smallest cluster. Ground states with dimer 
structures are possible in both elastic and floppy systems with 
$S=1/2$. The difference appears in their low-lying energy spectra. 
   An elastic quantum system has usually unique or finitely degenerate 
ground states.
   The lowest excitation mode is expected to have an energy gap and 
a dispersion both of the order of the typical exchange parameters. 
   On the other hand, unusual situations such as a macroscopic 
degeneracy of the dimer ground state \cite{takano} or dispersionless 
low energy excitations \cite{delta} occur in floppy systems. 
   These seem to be two different quantum manifestations of the 
classical floppiness.

   In this paper we report exact ground states with larger clusters than 
dimers in a frustrated AF chain. 
   The system contains a parameter 
which regulates the strength of the frustration, and can be either 
elastic or floppy according to the value of the parameter. 
   Wave functions with periodic cluster structures appear in the 
floppy regime. 

   The system is described by the Hamiltonian: 
$$
H=\sum_{i=1}^{N} h_{i} \eqno(1a)
$$
$$
h_{i} = J({\bf S}_{i}+{\bf S}_{i+1})\cdot 
({\bf T}_{i}^{(1)}+{\bf T}_{i}^{(2)})
+J'{\bf T}_{i}^{(1)}\cdot {\bf T}_{i}^{(2)} \eqno(1b)
$$
\noindent
where ${\bf S}_{i}$, ${\bf T}_{i}^{(1)}$ and ${\bf T}_{i}^{(2)}$ are spins 
with magnitude $S$ in the $i$-th unit cell. 
We assume that $J$ is positive.
The sub-Hamiltonian $h_{i}$ corresponds to a diamond-shape and so we 
call the total lattice shown in figure 1(a) simply a {\it diamond chain}. 

The system was studied earlier briefly by Sutherland and Shastry as an 
example of the superstability\cite{sutherland-s}. The case of $S$=1/2 
and $J'=2J$ was discussed as an example in the context of a general 
method to derive spin models with complete dimer ground 
states \cite{takano}.
The cases with $J'=\pm J$ were studied by Long et al \cite{long}. They 
numerically found the ground state with four-site clusters for $S=1/2$ 
and $J'=J$. 

Recently it was reported that the magnetic properties of a mixed 
molecular crystal of organic mono- and di-radicals are described 
by the Hamiltonian (1) for $S=1/2$ with $J\simeq 30$K and 
$J'\simeq -20$K\cite{exp}.

We treat the whole region of $J'$, though our main interest is in the 
frustrated region where $J'$ is positive. For a given $S$ the ratio 
$\lambda \equiv J'/J$ is the only parameter which determines the 
properties of the system. We assume periodic boundary conditions 
in the following, though most of the results do not depend on this 
assumption.

Before proceeding to the full quantum treatment, we examine the 
ground state spin configuration in the classical limit 
($S\rightarrow\infty$ with $JS^{2}$ finite). 
The ground state is given by a spin configuration which minimizes 
the energy of all the unit diamonds. Equation (1b) is rewritten as
\setcounter{equation}{1}
\begin{equation}
\label{eq2}
h_{i} = {J\over 2}\{ ({\bf T}_{i}+{\bf{\tilde S}}_{i})^{2} 
+ (\lambda - 1) {\bf T}_{i}^{2}-{\bf{\tilde S}}_{i}^{2} - 2\lambda S(S+1)\} 
\end{equation}
\noindent
by using ${\bf T}_{i}\equiv {\bf T}_{i}^{(1)}+{\bf T}_{i}^{(2)}$ and 
${\bf \tilde S}_{i}\equiv {\bf S}_{i}+{\bf S}_{i+1}$. In the classical 
limit $S(S+1)$ is replaced by $S^2$ in  (\ref{eq2}). For $\lambda<1$ it 
is seen that the energy is minimized when the following conditions are 
satisfied: $|{\bf T}_{i}|=2S$, $|{\bf \tilde S}_{i}|=2S$ and 
$|{\bf T}_{i}+{\bf\tilde S}_{i}|=0$. These conditions imply that 
${\bf T}_{i}^{(1)}$ and ${\bf T}_{i}^{(2)}$, and ${\bf S}_{i}$ and 
${\bf S}_{i+1}$ are parallel to each other, respectively, and further 
${\bf T}_{i}^{(j)}$ ($j$=1 or 2) and ${\bf S}_{i}$ are antiparallel to 
each other.
So this is a ferrimagnetic state where all the ${\bf S}_{i}$'s 
(${\bf T}_{i}^{(j)}$'s) are aligned parallel (antiparallel) to an axis. 
The ground state energy and the magnetization are given by 
$-(4-\lambda)JS^2$ and $S$ per unit diamond, respectively. 
The ground state is elastic in this region. 

   For $\lambda >1$, another expression of $h_i$ is useful, i.e. 
$h_{i}=(J\lambda /2)\{ ({\bf T}_{i}+{\bf\tilde S}_{i}/\lambda)^{2} 
-({\bf\tilde S}_{i}/\lambda)^{2}-2S(S+1)\}$. 
The energy minimum is realized if $|{\bf \tilde S}_{i}|=2S$ and 
$|{\bf T}_{i}+{\bf\tilde S}_{i}/\lambda |=0$. 
The former condition leads to all the ${\bf S}_{i}$'s being parallel 
to an axis. The latter condition is satisfied if and only if 
$\lambda\geq 1$, and means that ${\bf T}_{i}^{(1)}$ and 
${\bf T}_{i}^{(2)}$ form a triangle with ${\bf \tilde S}_{i}/\lambda$. 
They make an angle $\theta =\cos^{-1} 1/\lambda$ to the axis through 
${\bf S}_{i}$ and may be rotated about this axis simultaneously without 
raising the energy. The ground state contains $N$ degrees of freedom of 
free rotations and hence is floppy. The ground state energy is 
$-(\lambda +2/\lambda )JS^{2}$ per unit diamond. The ground state is 
again ferrimagnetic except for $\lambda=2$ with a magnetization 
$|2/\lambda -1|S$ per unit. 
Thus in the classical limit the diamond chain has two different ground 
state phases, the elastic one for $\lambda<1$ and the floppy one for 
$\lambda>1$.

We turn to the quantum case with general $S$. Hereafter we take an 
energy unit of $J=1$. We find easily from  (\ref{eq2}) that 
$[{\bf T}_{i}^{2},H]=0$; i.e., for all $i$, ${\bf T}_{i}^{2}=T_{i}(T_{i}+1)$ 
are good quantum numbers ($T_{i}=0, 1,\dots, 2S$).
With fixed $\{ T_{i}\}$,
the original problem of 3$N$ spins reduces to a problem of a linear 
chain with 2$N$ spins, where the ($2i-1$)-th site is occupied by the 
spin ${\bf S}_{i}$ and the 2$i$-th by ${\bf T}_{i}$. The energy of 
$J'$-bonds (i.e. a part of the energy proportional to $\lambda$) is 
determined solely by $\{ T_{i}\}$. It should be noted that if $T_{i}=0$ 
then there is no interaction between the left and the right side of 
${\bf T}_{i}$ and the whole lattice is decoupled. 

Let us first consider the eigenstates of an isolated unit diamond 
described by $h_{i}$.
In the lowest eigenstate of $h_{i}$ for a given $T_{i}$, the energy is 
\begin{equation}
\label{eq3}
E_{1}(T_{i})=T_{i}[{\lambda \over 2}(T_{i}+1) - (2S+1)] 
- \lambda S(S+1) 
\end{equation}
and the total spin is $2S-T_{i}$. Then we obtain, 

{\it Lemma 1.\/}\hspace{5 mm}
   For $T\geq 1$ and for $\lambda > \lambda_{\rm D}(T)$ where 
$\lambda_{\rm D}(T) \equiv 2(2S+1)/(T+1)$, it holds that 
$T_{i}\neq T$ for any $i$in the ground state of the diamond chain. 

{\it proof\/}\hspace{5 mm}
The total Hamiltonian is divided as $H=H'+h_{m}$, where $H'$ is the sum 
of $h_{i}$'s with $i\neq m$. Let us take a state $|\Psi _{0} \rangle$ 
whose wave function is given by the direct product of the ground state 
wave function of $H'$ and the singlet wave function of ${\bf T}_{m}$. 
Then $\langle \Psi_{0}|H|\Psi_{0} \rangle =E_{1}(0)+E'$, where $E'$ is 
the ground state energy of $H'$. Let $|\Psi \rangle$ be any state with 
$T_{m}=T\geq 1$. 
Then $\langle \Psi |H|\Psi \rangle = 
\langle \Psi |h_{m}|\Psi \rangle+\langle \Psi |H'|\Psi \rangle$. 
Clearly $\langle \Psi |h_{m}|\Psi \rangle \geq E_{1}(T)$ and 
$\langle \Psi|H'|\Psi \rangle\geq E'$;
also $E_{\rm 1}(T) > E_{\rm 1}(0)$ for $\lambda > \lambda_{\rm D}(T)$. 
Therefore $\langle \Psi |H|\Psi \rangle > \langle \Psi_{0}|H|\Psi_{0} 
\rangle$ and $|\Psi \rangle$ can not be the ground state of $H$. 
QED 

Since $\lambda_{\rm D}(T)$ decreases with $T$ from 
$\lambda_{\rm D}(1)=2S+1$ to $\lambda_{\rm D}(2S)=2$, we obtain 
the following result.

{\it Proposition 1.\/}\hspace{5 mm}
   For $\lambda > 2S+1 $, $T_{i}=0$ for all ${i}$ in the ground state. 
That is, all pairs of ${\bf T}_{i}^{(1)}$ and ${\bf T}_{i}^{(2)}$ form 
singlet dimers.

In this state all ${\bf S}_{i}$'s are decoupled from other spins and 
behave as free spins.
So we call this state a {\it dimer-monomer} (DM) state (figure 1(b)). 
Due to the free spins there is a $(2S+1)^{N}$-fold degeneracy in the 
DM state. 

   For $\lambda < 2S+1$, the DM state is not a ground state since one can 
lower the energy by introducing an isolated $T_{i}=1$ in the DM state. 
   For $S=1/2$ the ground state is composed of $T_{i}$'s with their 
magnitude 0 or 1. It is also true for $S\geq 1$ if 
$2S+1>\lambda>\lambda_{\rm D}(2)=2(2S+1)/3$. 
   For example, let us assume the configuration of $\{ T_{i}\}$ as 
$\{ 1010111011110110\} $ in the chain with 16 diamonds. 
Since zero ${T_{i}}$'s decouple the system to $n$-diamond clusters, 
the lowest energy for this configuration is simply given by 
$5E_{1}(0)+2E_{1}+E_{2}+E_{3}+E_{4}$, where $E_{n}(n\geq 1)$ is 
the lowest energy of a cluster with $n$ diamonds for the configuration 
with all $T_{i}$'s equal to 1.
Taking account of an $E_{1}(0)$ accompanying each $n$-diamond 
cluster, the average energy of this cluster per diamond measured from 
the DM state is given by $e_{n}=(E_{n}-nE_{1}(0))/(n+1)$. 
If the minimum value of $e_{n}$ occurs at $n=n_{\rm m}$, then the 
ground state in the thermodynamic limit ($N\rightarrow \infty$) is 
realized by a regular array of $n_{\rm m}$-clusters with isolated 
$T_{i}=0$'s between them.
A finite $N$ may cause a mismatch of the cluster size. 
If the minimum $e_n$ is realized at more than one value of $n$, 
then the ground state will have a macroscopic degeneracy. 
It seems plausible that the ground state is the state with 1-diamond 
clusters on every other diamonds for $\lambda$ less than but close 
to $2S+1$, since a simple variational argument leads to 
$E_{n} \ge nE_{1}$ and usually the difference between both sides of 
the inequality is fairly large. 
The wave function of this state is a direct product of those of four 
spin clusters ($T_{i}=1$) and dimers ($T_{i}=0$) as shown in figure 1(c) 
\cite{long}.
Therefore we call this state a {\it tetramer-dimer} (TD) 
state \cite{not-ss}. 

{\it Proposition 2\/}\hspace{5 mm}
There exists a finite region $\lambda_{\rm C1} < \lambda < 2S+1$ 
in which the TD state is the ground state of the diamond chain in 
the thermodynamic limit. 

{\it proof\/}\hspace{5 mm}
We show that $e_{n}$ is the minimum only at $n=1$ for a region 
$\lambda_{\rm C1} < \lambda < 2S+1$. We denote the ground state 
energy of a nearest neighbor AF linear chain with $n$+1 spins of 
magnitude $S$ on odd sites and $n$ spins of magnitude unity on even 
sites as ${\tilde E}_{n}$. 
Then we have $E_{n}-nE_{1}(0)={\tilde E}_{n}+n\lambda$. 
Let us first assume $\delta \equiv {\tilde E}_{2} - 2{\tilde E}_{1}>0$. 
If $n$ is even, the Hamiltonian for ${\tilde E}_{n}$ is divided into $n/2$ 
sub-Hamiltonians, each equivalent to that for ${\tilde E}_{2}$. 
A variational argument gives an upper bound on ${\tilde E}_{n}$ as 
$n{\tilde E}_{1}+n\delta /2$ for even $n$, and as 
$n{\tilde E}_{1}+(n-1)\delta /2$ for odd $n$ $(\geq 3)$. 
Since $e_{1}=[\lambda-(2S+1)]/2$, a lower bound on $e_{n}-e_{1}$ is 
given as $\frac{n-1}{2(n+1)} [\lambda -(2S+1)+\delta]$, which is 
positive for $\lambda >2S+1-\delta$ and $n \geq 2$. 
So we obtain $\lambda_{\rm C1} \leq 2S+1-\delta$. 

Next we assume that $\delta =0$ and deduce a contradiction. 
We divide the Hamiltonian for ${\tilde E}_{2}$ with five spins as 
${\tilde h}_{1} + {\tilde h}_{2}$, where each one is the Hamiltonian 
for three spins, two of magnitude $S$ and one of magnitude unity in 
the middle. A simple variational argument proves that $\delta \geq 0$ 
and the assumption implies that the ground state of 
${\tilde h}_{1} + {\tilde h}_{2}$ is also the ground state of 
${\tilde h}_{1}$. 
The Lieb-Mattis theorem\cite{lieb-m} holds in this case and implies 
that the total spin of the ground state is $|3S-2|$ and that for 
${\tilde h}_{1}$ is $2S-1$. The theorem also tells us that the ground 
state with the $z$-component of the total spin $S_{\rm T}^{z}$ 
contains all the $S^{z}$-diagonal states compatible with 
$S_{\rm T}^{z}$. The state $|S,0,S,-1,S-1 \rangle$ is contained in the 
ground state with $S_{\rm T}^{z}=3S-2$. 
But the fact that the total spin of three spins on the left must be 
$2S-1$ eliminates the presence of the above state in the ground state 
and proves the failure of the assumption. QED 

{\it Remark\/}\hspace{5 mm}
We can show that $\lambda_{\rm C1} \leq 1$ for $S=1/2$. 
The proof will be given elsewhere\cite{takano-ks}.

   For each $S$, we need numerical studies of finite size systems to 
determine the precise value of $\lambda_{\rm C1}$ and to find the 
ground state structure for $\lambda < \lambda_{\rm C1}$. 
The analysis for $S=1/2$ and 1 will be given later. 

   For negative or positive but small $\lambda$, we expect that the 
ground state is a ferrimagnetic state corresponding to the complete 
ferrimagnetic ground state in the classical limit. 
   For $\lambda \leq 0$ we have a general result.

{\it Proposition 3 \/}\hspace{5 mm}
   For $\lambda \leq 0$ the ground state is ferrimagnetic, i.e. it has both 
ferromagnetic and AF long-range order. 
   For all $i$, $T_{i}=2S$ in the ground state.

{\it proof \/}\hspace{5 mm}
We divide the total lattice into two sublattices, where ${\bf S}_{i}$'s 
are on the A sublattice and ${\bf T}_{i}^{(j)}$'s on the B sublattice. 
   For $\lambda \leq 0$ the Lieb-Mattis theorem\cite{lieb-m} implies 
that the total spin of the ground state is given by $NS$; i. e. 
ferromagnetic long range order (FLRO) exists. The positive 
(or negative) definiteness of the ground state wave function in terms 
of the $S^{z}$-diagonal basis implies that AFLRO is not less 
than FLRO\cite{shen-qt} and that $T_{i}=2S$ for all $i$. 
QED 

The ferrimagnetic ground state phase is expected to extend to positive 
$\lambda$ and we determine the phase boundary for $S=1/2$ by 
numerical methods. 

There is macroscopic degeneracy of the ground states at $\lambda 
=2S+1$. Since $E_{1}(1)=E_{1}(0)$ at this value of $\lambda$, any 
configuration of $\{ T_{i} \}$ with only $T_{i}=1$ or 0 and satisfying 
$\{T_{i}, T_{i+1}\}\neq \{1, 1\}$ for any $i$ gives ground states. 
We estimate the total degeneracy $D$ by using an asymptotic expansion 
for large $N$ as 
$ \log D \approx N\log [(2S+1)(\sqrt{1+4/\alpha}+1)/2]$ where 
$\alpha = (2S+1)^{2}/(4S-1)$. The ratio of the residual 
to the total entropy is 0.424 for
$S=1/2$ and decreases with $S$ to 1/3 for $S=\infty$. 
Macroscopic degeneracy of the ground state can occur at other 
phase boundaries. 

%
Let us inspect the case of $S=1/2$.
We estimate the minimum energy of a cluster composed of $n$ unit 
diamonds where all $T_{i}$'s are unity. 
   For this purpose, we numerically calculated the ground state energy 
${\tilde E}_{n}$ of finite linear chains with $n+1$ spin one-halves 
and $n$ spin unities alternatingly aligned. 
By employing the Lanczos technique we obtained the energy up to $n=7$. 
The obtained data of ${\tilde E}_{n}$ fit nicely to an assumed 
asymptotic formula
${\tilde E}_{n}/(n+1) \simeq {\tilde e}_{\infty} + a/(n+1)$ with 
${\tilde e}_{\infty} = -1.454$. We have estimated the values of 
$\lambda$ where the state with $n$-diamond clusters gives the same 
energy with the TD state by using the obtained ${\tilde E}_{n}$'s. 
The values are 0.763012, 0.819171, 0.847024, 0.862127, 0.871321, 
and 0.877464 for $n=2,3, \dots,7$.
   From the monotonically increasing behavior of these values, 
and the linear dependence of $e_{n}$ on $\lambda$ we conclude that 
the TD phase changes directly to the ferrimagnetic phase where 
$T_{i}=1$ for all $i$.
The critical $\lambda$ is estimated as 0.909 from the value of 
${\tilde e}_{\infty}$.
Thus we have completely determined the ground state phase diagram 
of the diamond chain for $S=1/2$.
It consists of three phases; the DM phase for $\lambda > 2$, the 
TD phase for $0.909 < \lambda < 2$ and the ferrimagnetic phase for 
$\lambda < 0.909$ (figure 2). 

In the case of $S=1$ we have obtained ground states with clusters 
larger than tetramers. 
The ground state changes its character successively from the DM state 
to the TD state at $\lambda=3$, then to the heptamer ($n$=2)-dimer 
state at $\lambda=2.660$, to the state with 3-diamond clusters at 
$\lambda=2.583$ and suddenly to the state where $T_{i}=1$ for all $i$ 
(i.e. $n=\infty$) at $\lambda =2.577$. 
Therefore the ground state phase diagram has at least six phases. 
The detail will be reported elsewhere\cite{takano-ks}. 

Above we have shown that the spin system on the diamond chain have 
several ground state phases with cluster structures\cite{double2}. 
There appear clusters larger than conventional dimers. 
   For $S=1/2$ the TD state is especially interesting since it is 
nonmagnetic. We believe that these cluster
structures are quantum manifestations of the classical floppiness. 
The results might give some insight into the ground state of other 
floppy systems such as the kagom\' e antiferromagnet. 

We are grateful to T. A. Kaplan for careful reading of the manuscript 
and many helpful suggestions.
We thank T. Nakamura, Y. Okabe, S. Takada, S. D. Mahanti, 
J. B. Borysowicz and M. F. Thorpe for discussion and useful advice. 
We are indebted to T. Tonegawa for advice on numerical calculations.

\noindent
   Figure Captions
\medskip

\noindent
   Figure 1.
(a) The diamond chain. The circle represents a spin with magnitude 
$S$ and the solid (dashed) line represents exchange parameter $J (J')$.
(b) The dimer-monomer (DM) state. The unshaded oval represents a 
dimer ($T_{i}=0$). There are free spins (monomers) on the sites not 
closed by ovals.
(c) The tetramer-dimer (TM) state. The shaded oval represents a triplet 
pair ($T_{i}=1$) and the closed loop including four spins represents a 
tetramer.
\medskip

\noindent
   Figure 2.
Phase diagram for $S=1/2$ in the parameter space of $\lambda$ $(=J'/J)$.

\begin{references}
\bibitem{not-ss}
It was claimed in  \cite{sutherland-s} that the dimer-monomer state 
is stable for $\lambda >2$ for general $S$. It is only valid for $S=1/2$ 
and does not hold for $S\ge 1$ as proposition 2 says.
\bibitem{double2}
Similar arguments have been done for an S=1/2 double chain model 
which includes a parameter [6].  However, the model has no ground 
state with clusters larger than dimers for all parameter values.
\end{references}
\end{document}